# Upconversion channels in $Er^{3+}$:ZBLALiP: multicolour microspherical light source

**D G O'Shea[1,2], J M Ward[2,3], B J Shortt[2,3], M Mortier[4], P Féron[5] and S Nic Chormaic[1,2]**

[1]Physics Department, National University of Ireland, University College Cork, Ireland
[2]Tyndall National Institute, Prospect Row, Cork, Ireland
[3]Department of Applied Physics and Instrumentation, Cork Institute of Technology, Bishopstown, Cork, Ireland
[4]Laboratoire de Chimie de la Matière Condensée de Paris, ENSCP, 11 rue Pierre et Marie Curie, 75005 Paris, France
[5] ENSSAT-FOTON (CNRS-UMR 6082) Université de Rennes 1, 6 rue de Kerampont, BP 80518, 22305 Lannion Cedex, France

**Abstract**

We present results on the realization of an all-taper coupled, multicolour microspherical glass light source fabricated from the erbium doped fluoride glass ZBLALiP. Whispering gallery mode lasing and upconversion processes give rise to laser and fluorescent emissions at multiple wavelengths from the ultraviolet to the infrared. A single-mode tapered fibre coupling scheme, which serves to both launch pump light at 980 nm into the microresonator cavity and collect the resulting infrared lasing, is employed. We demonstrate the use of a second separate fibre taper to outcouple the upconversion spectrum over several hundred nanometers. Thirteen discrete emissions ranging from 320 to 850 nm have been observed in the upconversion spectrum. The absorption and fluorescence properties are studied and the processes responsible for the generation of the observed wavelengths are outlined.

## 1. Introduction

In recent years, much effort has been devoted to the exploitation of upconversion mechanisms in rare-earth ion doped materials to yield a range of wavelengths from near infrared to ultraviolet. The spectral properties of the resulting fluorescence and/or lasing emissions depends on the choice of dopant and the host matrix in which it is embedded. The addition of multiple dopants can increase the range of spectral emission [1] while simultaneously reducing the lasing thresholds of devices fabricated from these materials [2]. One such class of device which has interesting optical properties is the rare-earth doped microspherical whispering gallery mode laser [3,4]. These miniature resonant cavities are appealing for applications in laser engineering due to their small size of several tens of microns in diameter, high-Q whispering gallery modes reaching values of $10^{10}$, and small mode volumes [5]. The close proximity of numerous energy levels in triply-ionized rare earth ions is advantageous for obtaining fluorescent and lasing emissions through various upconversion mechanisms such as excited state absorption (ESA), energy transfer upconversion (ETU) and photon avalanche (PA) [6]. This combination of solid state active medium and resonant cavity has resulted in much research focusing on the characterization of doped microspheres.

Three-photon upconversion lasing yielding red and blue emissions about 480 and 800 nm in thulium-doped fluorozirconate (ZBLAN) glass microspheres has been reported in ref. [7]. The same authors also observed red and green upconversion lasing in erbium doped ZBLAN glass microspheres [8]. In both cases 1064 nm free-space laser pulses focused on the microsphere surface acted as the pump source. The first microspherical laser pumped using a fibre taper was demonstrated when an $Yb^{3+}$:$Er^{3+}$ -doped phosphate glass microsphere was pumped at 980 nm [9]. The microsphere lased in the infrared lasing about 1550 nm with a threshold power of 60 $\mu$W, and emitted green fluorescence which was attributed to a two-photon ESA process. Other authors [10,11] have demonstrated similar green-emitting microspheres fabricated from erbium-doped ZBLAN and titanium barium (TBS) glasses using prism coupling and free-space laser pumping schemes respectively. Several groups [3,4,12,13] have also made considerable progress in the development of microspherical lasers operating in the infrared region, in particular with emission wavelengths falling in the C- and L-bands.

The vast majority of work on activated glasses found in the literature deals with bulk samples, waveguides, and to a lesser extent, microcavities. Traditionally the fluorescence from these structures is collected by free-space coupling of the scattering into a spectrometer. It has been known for some time [14] that it is not possible to use SMF28 to outcouple all lasing and fluorescence emissions from a microcavity due the narrow transparency window of the fibre. Furthermore, fabrication of adiabatic multimode fibre tapers is not possible, making pumping and light collection unrealistic [15].

We report here on the realization of an all-taper coupled, multiwavelength microspherical light source fabricated from the novel erbium-doped fluoride glass ZBLALiP [4]. The suitability of fluoride glasses as rare-earth host materials has long been known [16] as their low phonon energies determine multiphonon relaxation rates and, ultimately, the efficiency of upconversion processes. Upconversion emissions from the ultraviolet to the infrared are observed and outcoupled and we report, to the best of our knowledge, the first observation of 320 nm upconversion emission following 980 nm pumping of $Er^{3+}$ and the first use of a fibre taper to outcouple the upconversion spectrum over several hundred nanometers. The optical absorption, infrared lasing and fluorescence properties are investigated.

## 2. Experimental

### *2.1. ZBLALiP material properties*

A detailed study of some fundamental material properties of $Er^{3+}$ doped ZBLALiP ($ZrF_4$-$BaF_2$-$LaF_3$-$AlF_3$-LiF-$PbF_2$) glass has been presented in an earlier paper [4]. That work described the improved stability of ZBLALiP with respect to similar fluorozirconate glasses and its suitability as a host for rare-earth ions. Additional material properties are presented here, including the absorption spectrum for a 5.8 mm thick bulk sample of 0.2 mol% $Er^{3+}$:ZBLALiP (c.f. figure 1). Twelve absorption bands are indicated in the figure and represent the transitions from the ground state $^4I_{15/2}\rightarrow A_N$ with $A_1={}^2G_{9/2}+{}^2K_{15/2}+{}^2G_{7/2}$, $A_2={}^4G_{11/2}$,

$A_3={}^2H_{9/2}$, $A_4={}^4F_{3/2}$, $A_5={}^4F_{5/2}$, $A_6={}^4F_{7/2}$, $A_7={}^2H_{11/2}$, $A_8={}^4S_{3/2}$, $A_9={}^4F_{9/2}$, $A_{10}={}^4I_{9/2}$, $A_{11}={}^4I_{11/2}$ and $A_{12}={}^4I_{13/2}$. We use the same glass sample as described in ref. [4].

Figure 2 presents the absorption and emission cross-section measurement for 0.2 mol% $Er^{3+}$:ZBLALiP in the C-band. Emission spectra for both bulk and powder ZBLALiP samples are shown. The absorption spectra were recorded using a double-beam Cary 17 spectrometer (resolution ~0.1 nm). The refractive index, *p*, of ZBLALiP is 1.49 at 1550 nm.

*2.2. ZBLALiP microspheres*

The $Er^{3+}$:ZBLALiP microspheres studied for this work contain erbium at a concentration of 0.2 mol%, which equates to $4\times10^{19}$ $Er^{3+}$ ions/cm$^3$. It should be noted that, at this concentration, energy transfer (ET) is not expected to play a major role and, as such, excited state absorption (ESA) is the dominant mechanism in the upconversion process [17]. Nevertheless, due to the non-trivial problem of distinguishing the two mechanisms, some authors have suggested that ET may play a role even when the dopant concentration is below the critical value of about 1 mol% [18,19,20].

In our work, the C-band lasing characteristics and upconversion spectrum are investigated at room temperature using a single mode 980 nm laser diode pump with a spectral width of about 1 nm and under CW pumping conditions. A detailed description of the experimental approach is given elsewhere [21]. To achieve efficient coupling of the pump into the microsphere and efficient collection of the infrared lasing spectrum, a 1550 nm SMF-28 fibre was adiabatically tapered to 1 $\mu$m diameter using a direct heating technique [22]. The transmission loss of these tapers is typically less than 0.1 dB/cm. Optimum mechanical alignment of the taper and microsphere is achieved by moving the microsphere to different locations along the taper until the microsphere emission spectrum about 1550 nm is maximized. This spectrum is recorded by connecting one end of the pump fibre taper to an optical spectrum analyzer. The transmission through the pump fibre is also monitored during the alignment and typically 10–15 % of the pump light is coupled into the microsphere. This coupling efficiency is, in part, due to the broad spectral nature of the pump laser relative to the whispering gallery mode resonances of the microsphere and can be improved by the use of a narrow linewidth pump source [23]. Characteristic whispering gallery mode bands about the sphere equator are also observed.

The upconversion spectrum is detected in two ways. The first is via the use of a second fibre half-taper located on the microsphere equator diametrically opposed to the pump fibre taper, as depicted in the inset to figure 3. This detection fibre taper is non-adiabatic, fabricated from multimode fibre (Thorlabs AFS50/125Y) and has a diameter of 2 $\mu$m at the point of contact with the microsphere. It is connected to an optical spectrometer with 0.4 nm spectral resolution. We employ the two-taper system as it enables us to simultaneously detect the lasing emissions about 1550 nm via the pump taper and the upconversion emissions via the second multimode fibre taper. This technique allows us to study the competition between both sets of processes. The second method relies on the use of a high numerical aperture fibre coupled spectrometer to detect the free-space scattering spectrum from the microsphere.

## 3. Results and discussion

### *3.1. Judd-Ofelt analysis*

From the absorption spectrum, important information can be assessed using the absorption cross-section expression given as [19]

$$\sigma_{abs} = \frac{-\ln(10)OD}{lN'} \tag{1}$$

where *OD* is the optical density, *l* is the sample thickness, and $N'$ is the $Er^{3+}$ concentration of $3.76 \times 10^{19}$ $cm^{-3}$. The Judd-Ofelt (JO) intensity parameters $\Omega_t$ (t = 2,4,6) are derived from the electric-dipole contributions of the measured line strengths by a least-squares-fitting approach to minimize the rms deviation with the calculated line strength. The JO intensity parameters, $\Omega_2 = 3.12 \pm 0.15 \times 10^{-20}$ $cm^2$, $\Omega_4 = 1.49 \pm 0.07 \times 10^{-20}$ $cm^2$, and $\Omega_6 = 1.15 \pm 0.03 \times 10^{-20}$ $cm^2$ as calculated in ref. [4] are used throughout our calculations. A spectroscopic quality factor can be defined as $\Omega_4 / \Omega_6$, which gives 1.30 for these intensity parameters. These JO parameters are slightly higher than those for $Er^{3+}$:ZBLAN ($\Omega_2 =$

2.37 × $10^{-20}$ $cm^2$, $\Omega_4$ = 1.24 × $10^{-20}$ $cm^2$, and $\Omega_6$ = 0.73 × $10^{-20}$ $cm^2$) [24]. A higher $\Omega_2$ indicates greater asymmetry of the host glass, while a higher $\Omega_6$ has a dominant influence on the electric dipole line strength, $S^{ed}$, of the lasing transition $^4I_{13/2} \rightarrow {}^4I_{15/2}$, as shown by Weber *et al.* [25] in the expression

$$S^{ed}\left[{}^4I_{13/2} \rightarrow {}^4I_{15/2}\right] = 0.0195\Omega_2 + 0.1173\Omega_4 + 1.4316\Omega_6, \quad (2)$$

Equation (2) can be expressed more generally as

$$S^{ed} = \sum_{t=2,4,6} \Omega_t \left| \left\langle (S,L)J \left\| U^{(t)} \right\| (S',L')J' \right\rangle \right|^2, \quad (3)$$

where $\left| \left\langle (S,L)J \left\| U^{(t)} \right\| (S',L')J' \right\rangle \right|^2$ are the double reduced matrix elements of the unit tensor operator $U$ and are assumed to be independent of the host material [25]. Therefore, we have taken the matrix elements from refs. [25, 26] for our calculations and these are reproduced in table 1.

The magnetic dipole line strengths are calculated from

$$S^{md} = \frac{1}{4m^2c^2} \left| \langle (S,L)J \rangle \| L + 2S \| (S',L')J' \right|^2, \quad (4)$$

where $\left| \langle (S,L)J \rangle \| L + 2S \| (S',L')J' \right|^2$ are the double reduced matrix elements of the $L$+$2S$ operator, where $L$ is the total orbital angular momentum and $S$ is the total spin.

The data yielded from equations (1) to (4) combined with the JO intensity parameters, are used to calculate important radiative properties. The total spontaneous radiative transition probability or rate between level $J$ and a lower level $J'$, $A_{JJ'}$, is given by

$$A_{JJ'} = A_{JJ'}^{ed} + A_{JJ'}^{md} = \frac{64\pi^4}{3h\lambda^3(2J+1)} \left( \chi_{ed} S^{ed} + \chi_{md} S^{md} \right), \quad (5)$$

where $\lambda$ is the mean wavelength of the absorption, $\chi_{ed}$ is the local field correction factor for electric dipole transitions, $\chi_{ed} = p(p^2+2)^2/9$ and $\chi_{md}$ is the local field correction factor for magnetic dipole transitions $\chi_{md} = p^3$. In this work, $A_{JJ'}^{md}$ is calculated using the values given in ref. [25] for $LaF_3$ ($A'^{md}_{JJ'}$) corrected for the refractive index difference using the relation $A^{md}_{JJ'} = (p/p')^3 A'^{md}_{JJ'}$ where $p'$ ($p$) is the refractive index of $LaF_3$ (ZBLALiP). The radiative lifetime, $\tau_R$, of transitions between levels $J$ and $J'$ is related to the inverse of $A_{JJ'}$ by

$$\tau_R = \frac{1}{\sum_{J'} A_{JJ'}}, \quad (6)$$

and the branching ratio, $\beta$, between $J$ and $J'$ is simply the ratio of $A_{JJ'}$ for the initial level to the sum of $A_{JJ'}$ for all the lower levels given by

$$\beta = A_{JJ'}\tau_R, \quad (7)$$

Table 1 shows the radiative probabilities, electric dipole line strengths, branching ratios, and radiative lifetimes for the transitions. Several levels have high branching ratios for transitions to the ground level and intermediate levels, many of which have been observed in the emission spectrum in figure 4. For example, the energy level diagram in figure 5 attributes the 550 nm emission to ${}^4S_{3/2}\rightarrow{}^4I_{15/2}$ ($\beta = 0.66$) and 849 nm emission to ${}^4S_{3/2}\rightarrow{}^4I_{13/2}$ ($\beta = 0.28$); however, the observed intensity ratio disagrees with that predicted from the branching ratios. Three emissions (320 nm, 403 nm, and 470 nm) are attributed to the transitions ${}^2P_{3/2}\rightarrow{}^4I_{15/2}$ ($\beta = 0.12$), ${}^2P_{3/2}\rightarrow{}^4I_{13/2}$ ($\beta = 0.41$), and ${}^2P_{3/2}\rightarrow{}^4I_{11/2}$ ($\beta = 0.53$), respectively. The branching ratios predict both the 403 nm and 470 nm emissions to have similar intensities and the 320 nm emission to have a much weaker intensity according to the ratio 1 (320 nm): 3.4 (403 nm): 4.2 (470 nm), and the observed ratio is in reasonable qualitative agreement with ratios of 1:12.9:12.0, respectively.

The ${}^4F_{3,5/2}$ levels have unusually long lifetimes of 1.45 ms and 1.15 ms compared to other glasses such as $Er^{3+}$ doped fluoroindate glass which has a lifetime of 0.50 ms [27]. As can be seen from the energy level diagram, this is critically important for the UV and violet emissions from the ${}^2P_{3/2}$ level because it facilitates the third ESA, ${}^4F_{3,5/2}\rightarrow{}^2K_{13/2}$, by maintaining a large ${}^4F_{3,5/2}$ population. In $Er^{3+}$ doped fluoroindate glass, nonradiative decay rapidly depletes the $^4F_{3,5/2}$ population, effectively preventing any further ESA.

### *3.2. McCumber theory*

The emission cross-section, $\sigma_{ems}$, can be calculated from the absorption spectrum using McCumber theory under the assumption of a strongly phonon coupled system [28],

$$\sigma_{ems}(\lambda) = \sigma_{abs}(\lambda)\frac{Z_L}{Z_U}\exp\left(\frac{E_{UL}}{k_B T}\right)\exp\left(-\frac{hc}{\lambda k_B T}\right), \tag{8}$$

where $Z_L$ ($Z_U$) is the partition function of the lower (upper) state given by a summation of continuous levels $\sum_i \exp(\Delta E_i / k_B T)$, where $\Delta E_i$ is the energy difference between the Stark split sublevels and the lowest energy level in the lower (upper) manifold, $E_{UL}$ is the energy difference between the lowest sublevels in the upper and lower manifolds, $h$ is Planck's constant, $c$ is the speed of light, $k_B$ is Boltzmann's constant, and $T$ is temperature in Kelvin. Bulk $Er^{3+}$:ZBLALiP has a peak emission cross-section of $4.6 \times 10^{-21}$ cm$^2$ in the C-band, which compares well with $4.9 \times 10^{-21}$ cm$^2$ measured for fluoroindate glasses [27].

### *3.3. Upconversion fluorescence and 1.5 μm lasing*

Successful excitation of the $^4I_{13/2} \rightarrow {}^4I_{15/2}$ transition results in whispering gallery mode lasing resonances within the microsphere, as shown in figure 3. The microsphere eccentricity is determined from $\varepsilon = (\Delta\omega_{ecc}/\omega_{nml}) \cdot \left(l^2/|m| - 1/2\right)$, where $l$ and $m$ are the angular mode numbers, and $\omega_{nml}$ is the angular frequency of the mode. This is calculated to be ~2.1 % for $l = |m|$ with corresponding azimuthal mode splitting, $\Delta\omega_{ecc}$, of ~11 GHz. Analysis of the lasing peaks enables us to measure the free spectral range (FSR), $\nu_{FSR}$, between successive mode groups with a difference in angular mode number $l$ of 1, to be ~7.8 nm (0.97 THz). This agrees closely with the calculated value of $\nu_{FSR} = c/\pi N D$ ~7.9 nm (0.96 THz), indicating a microsphere diameter of $D \approx 66$ μm as expected. It can be seen that each group of modes contains four distinct lasing peaks which are due to the excitation of multiple radial modes in the sphere. The peak power of the lasing peaks presented here is ~30 nW, although we have also observed lasing in a single mode with a peak value of ~500 nW. The spacing between TE and TM modes ($\nu_{TETM}$) of 6.8 nm in figure 3 agrees well with the calculated value of 6.4 nm. The loaded cavity Q is typically 2–$5\times10^4$ and is measured as the ratio of peak FWHM to resonance wavelength.

The $Er^{3+}$:ZBLALiP upconversion spectrum recorded coincidently to the C-band lasing spectrum (figure 3) is shown in figure 4. This spectrum is produced by splicing two separate spectra together for both scattering and free-space detection schemes. This was necessary due to the limited dynamic linear range of the optical spectrometer used and in order to maximize the signal-to-noise ratio of the lower intensity transitions. The spectra about the green were recorded for a detector integration time reduced by a factor of 40. Thirteen discrete peaks have been noted in the scattering spectrum (figure 4(a)) and are numbered from T1-T13. Approximately 700 $\mu$W of 980 nm pump light was coupled into the microsphere. The detection efficiency of the upconversion outcoupling taper has not been optimized, and this, combined with increased fibre attenuation towards wavelengths shorter than about 500 nm, results in transitions T1 and T5 not being resolved in the taper spectrum (figure 4(b)). The taper and scattering spectra have been corrected for the responsivity of the spectrometer detector as a function of wavelength. The total power outcoupled and detected in the taper spectrum is 27 pW for the green transitions (T6 and T7) and 7 pW for T13, with corresponding values for the scattering spectrum of 0.5 nW and 0.1 nW respectively.

The transitions have initially been identified on the basis of the erbium energy level diagram shown in figure 5, the JO analysis in table 1, and on the accepted $Er^{3+}$ emission wavelengths in the literature. The centre wavelength associated with each transition is shown in figure 5. Further understanding of the transition(s) responsible for a particular wavelength can be obtained by the intensity dependence behaviour of the fluorescence at a particular wavelength. In the case of ESA, the dependence of the fluorescence intensity upon pump power often exhibits a simple power law dependence with slope $p$ equal to the number of photons absorbed. Measurements of this dependence for the case of the green transitions T6 ($^2H_{11/2}\rightarrow{}^4I_{15/2}$) and T7 ($^4S_{3/2}\rightarrow{}^4I_{15/2}$) are presented in figure 6. A slope corresponding to $n$ = 1.58 for T6 and $n$ = 1.66 for T7 has been recorded whereas a slope of 2 is expected. This can be understood as depletion of the $^4I_{15/2}$ ground state [7]. The limited resolution of the spectrometer used to study the upconversion combined with the smaller microsphere FSR at shorter wavelengths means that the whispering gallery mode structure is only visible for emissions at wavelengths longer than about 800 nm. The mode structure for the 849 nm (T13) emission is distinguishable in the inset of figure 4(a).

Our results demonstrate a broad range of emission from 320 nm to 849 nm. We suggest that the combination of modest cavity quality factor (typically $5\times10^4$) and low whispering gallery mode volume (approximately 2000 $\mu m^3$) serves to significantly enhance the probability of excited state absorption by enhancing the pump field strength (~$10^3$ V/m per photon) within the microsphere. This is reflected in the fact that a large number of the observed transitions are likely the result of 3-photon and 4-photon absorption events for what is a relatively modest pump power (sub mW) coupled into the microsphere. The emission line centred at 320 nm (T1) is most likely a 4-photon process populating the $^2P_{3/2}$ state, which in turn radiatively relaxes to the $^4I_{15/2}$ ground state. Three pump photons populate the $^2G_{7/2}$ state. The $^2G_{7/2}$–$^2P_{3/2}$ energy gap of ~3600 cm$^{-1}$ combined with a phonon energy of 650 cm$^{-1}$ is such that the $^2P_{3/2}$ state is unlikely to be populated by energy transfer processes such as thermalization and, as such, we suggest the absorption of a fourth pump photon is necessary to explain T1. This transition has previously been reported in $Er^{3+}$-doped glass, but was identified as a 3-photon process following pumping at 637 nm [29]. The same authors also reported emissions corresponding to the transitions T3, T4, T7 and T9. Transition T1 has also been reported [30] in an $Yb^{3+}$:$Er^{3+}$ co-doped material although in that case the excitation process involved successive energy transfers from $Yb^{3+}$. It should be noted that this is very different from our work which involves a single dopant, does not rely on a sensitizer ion, and, ETU does not play a major role due to the low dopant concentration. The transitions $^2P_{3/2}\rightarrow{}^4I_{13/2}$ (T3) and $^2P_{3/2}\rightarrow{}^4I_{11/2}$ (T4) are also thought to involve 4-photon excitation. Transitions T1, T3 and T4 have been observed following pumping at 545 nm [31]. Transitions T6, T7, T12 and T13 have been reported following 980 nm pumping [32]. The ultraviolet line at 383 nm (T2) and the 492 nm emission (T5) have been reported under 973 nm excitation [33]. $Er^{3+}$ emission at 700 nm has been attributed to the transition $^4F_{7/2}\rightarrow{}^4I_{13/2}$ following absorption of three pump photons at 1480 nm in ref. [34]. We presume this assignment was on the basis of the 3-photon resonance $^4F_{7/2}\rightarrow{}^4I_{15/2}$. We have also observed emission about 700 nm (T11) but propose it to follow the transition $^2H_{9/2}\rightarrow{}^4I_{11/2}$ based on the observed intensity ratio compared with the branching ratios calculated from JO theory. Finally, we have observed emissions centred at 618 nm (T8) and 667 nm (T10). We are presently unaware of other reports on such emissions from $Er^{3+}$.

## 4. Conclusions

In conclusion, we have demonstrated a multiwavelength microsphere light source for microphotonic applications [35]. The $Er^{3+}$ transitions involved have been identified. We have used tapered optical fibres to both pump the microsphere and collect the resulting upconversion and C-band lasing emissions. We believe this to be the first use of fibre tapers to outcouple such a broad range of wavelengths from a microspherical light source. This device could find many uses as a miniature all-fibre-coupled, multicolour light source with, for example, application to evanescent wave sensing based on excitation of transitions in target molecules adsorbed on to the microsphere surface. Our results for the violet emissions in $Er^{3+}$:ZBLALiP indicate an attractive alternative to $Er^{3+}$:ZBLAN for information-storage applications where even very small powers are sufficient for data retrieval operations [29]. Future work will focus on improving the outcoupling efficiency of the device using waveguide technology for integrated optics [36].

## Acknowledgements

This work is funded by Science Foundation Ireland under grant 02/IN1/128. D. O'Shea acknowledges support from IRCSET through the Embark Initiative RS/2005/156. B. Shortt's e-mail address is brian.shortt@cit.ie.

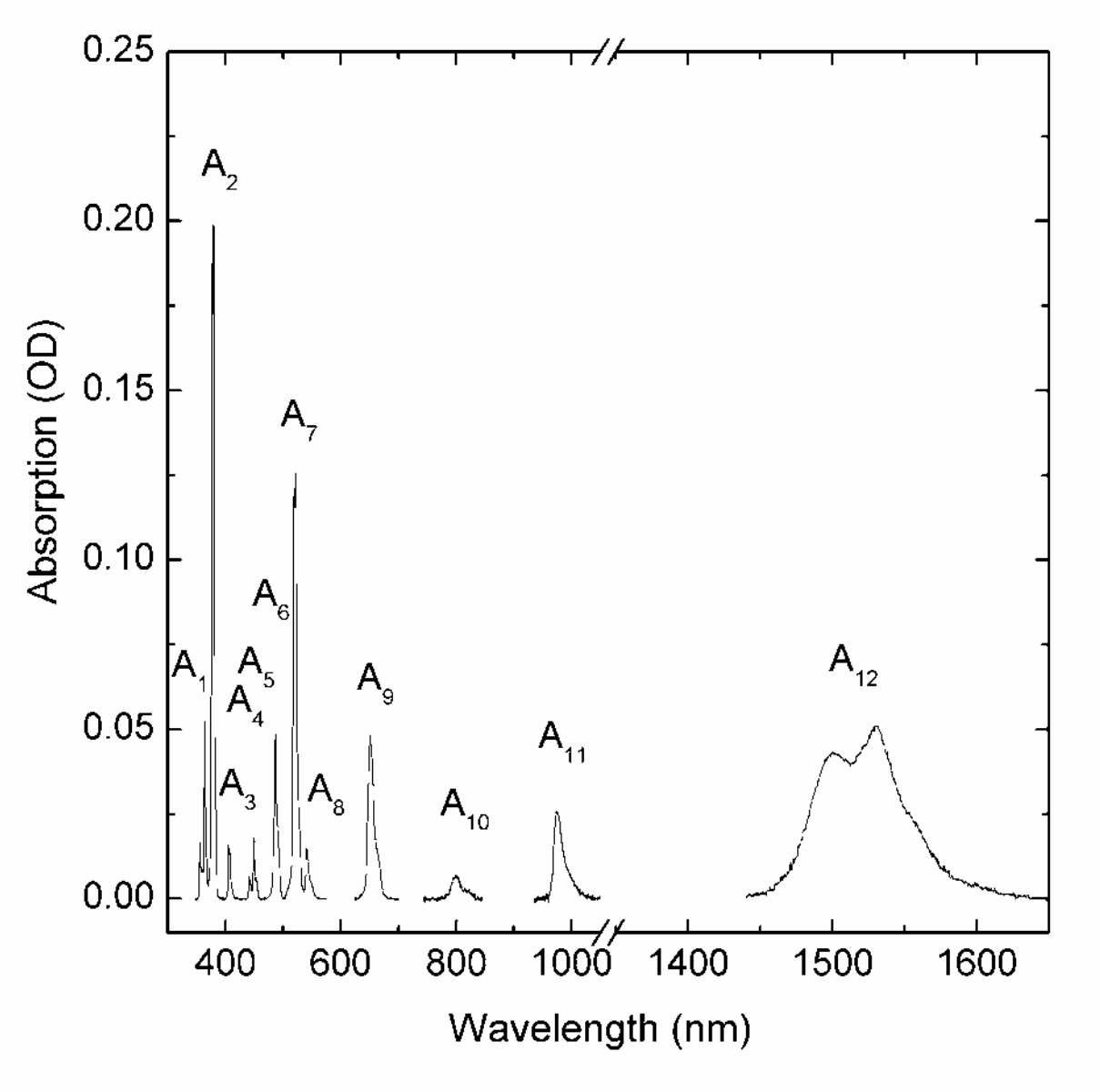

**Figure 1.** Absorption spectrum of 0.2 mol% $Er^{3+}$ doped ZBLALiP bulk glass.

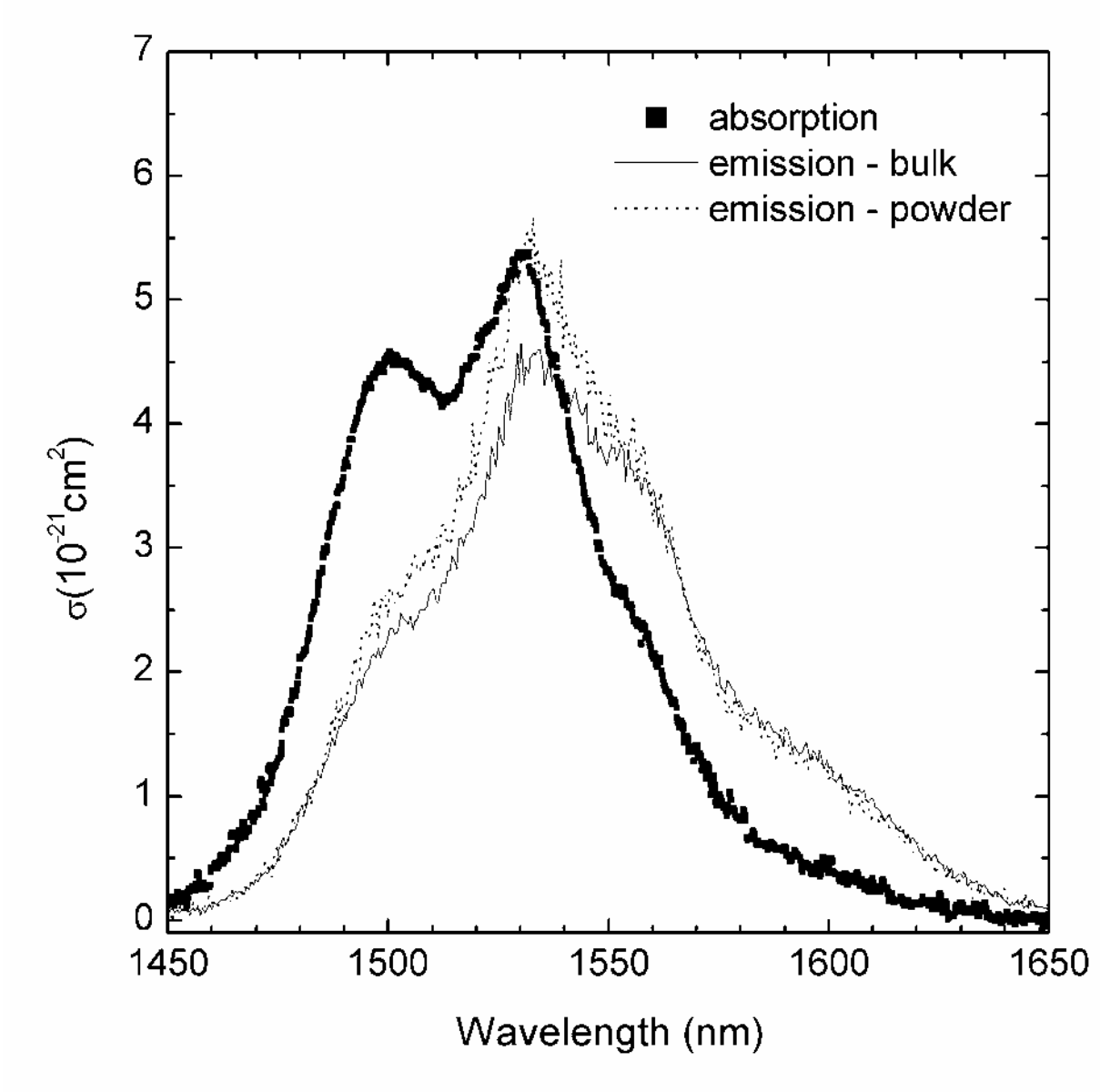

**Figure 2.** Absorption (bulk) and emission (bulk and powder) cross section measurements for 0.2 mol% $Er^{3+}$ doped ZBLALiP.

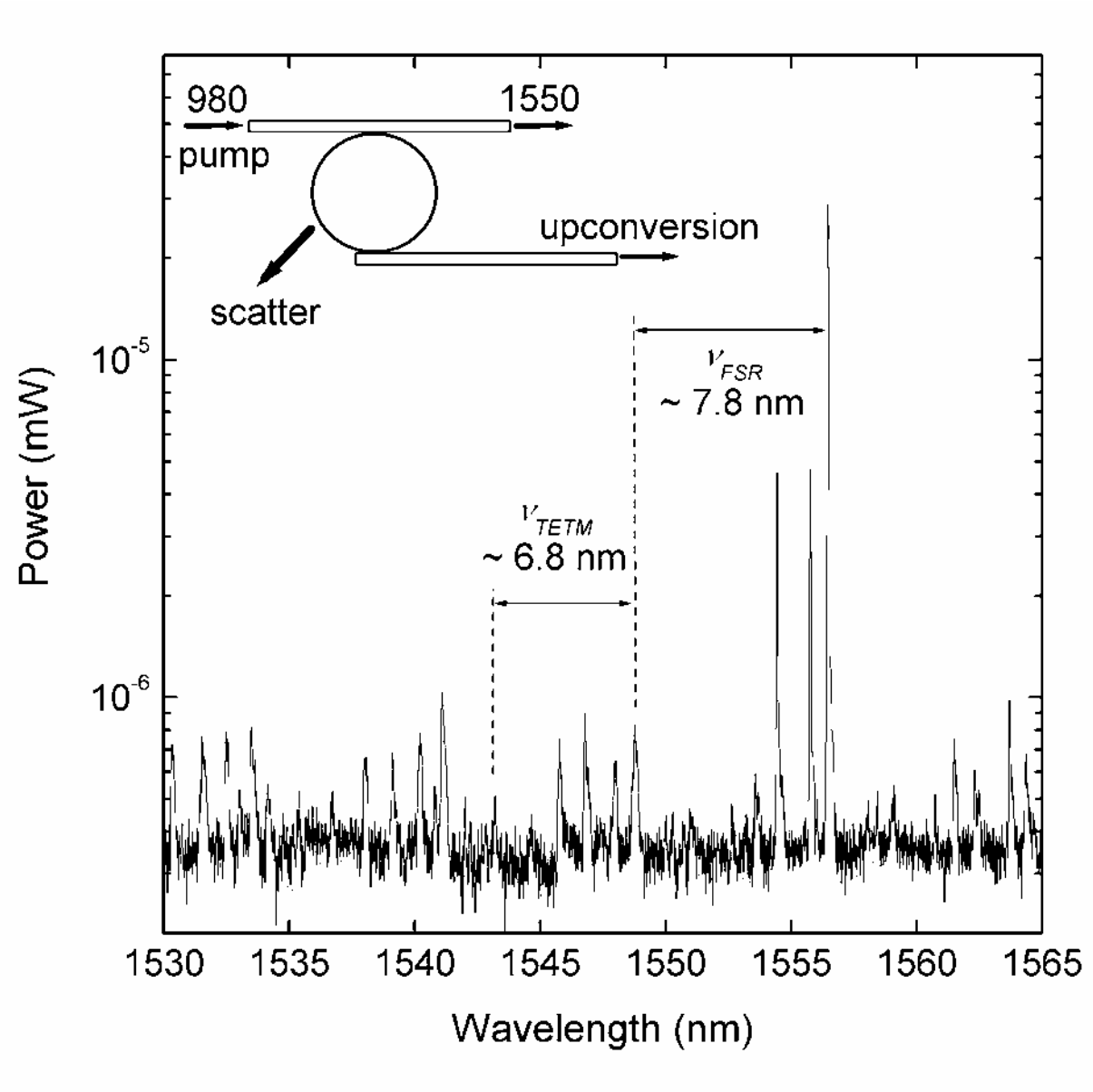

**Figure 3.** 66 $\mu$m diameter microsphere lasing spectrum about 1550 nm showing whispering gallery mode structure. Inset: microsphere with pump full taper (bottom) and upconversion outcoupling half taper (top).

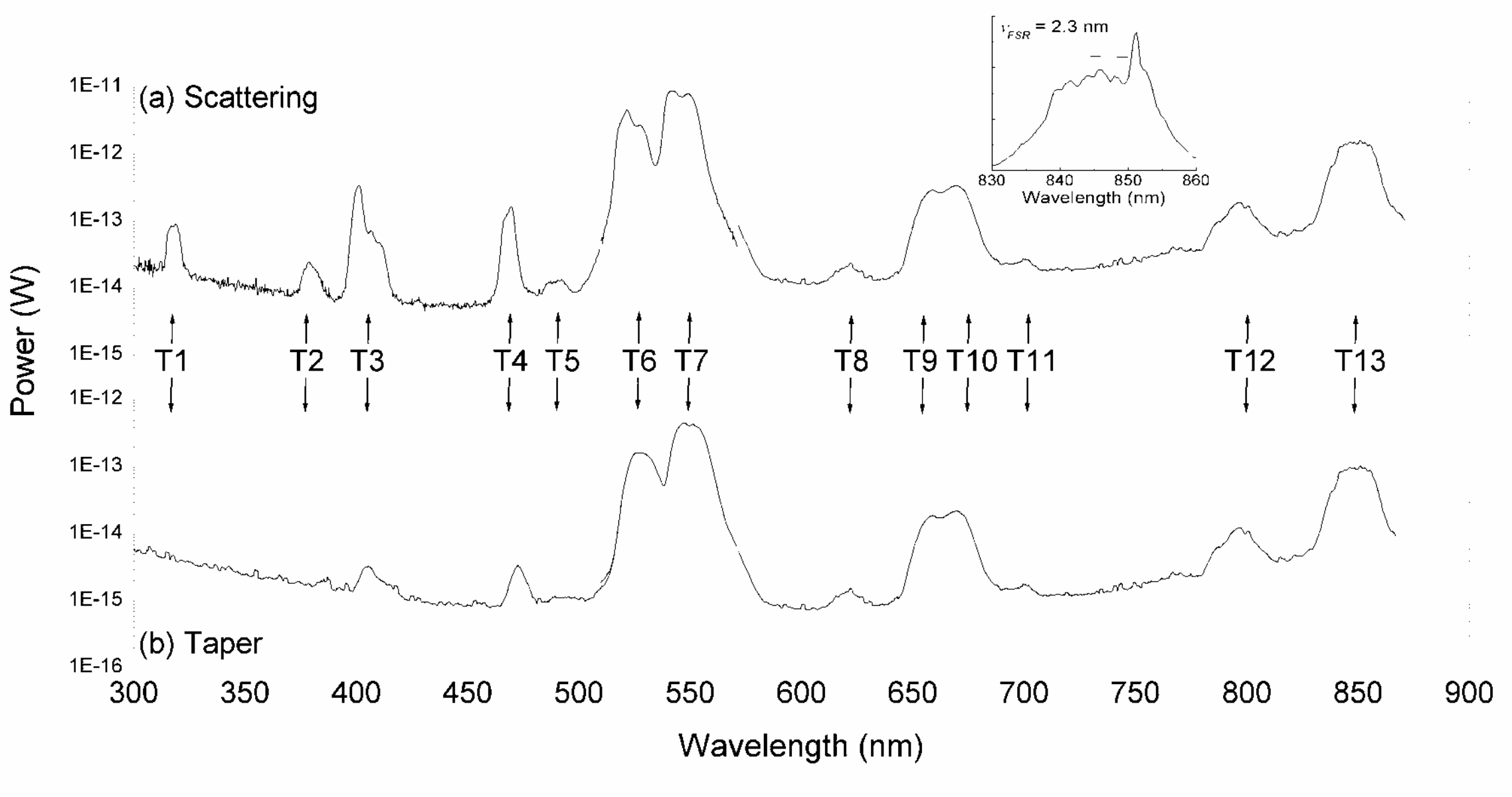

**Figure 4.** $Er^{3+}$:ZBLALiP microsphere upconversion spectrum with (a) scattering and (b) fibre taper detection schemes. The spectrometer integration time was reduced by a factor of 40 about the green transitions T6 and T7. Inset: whispering gallery mode structure about T13.

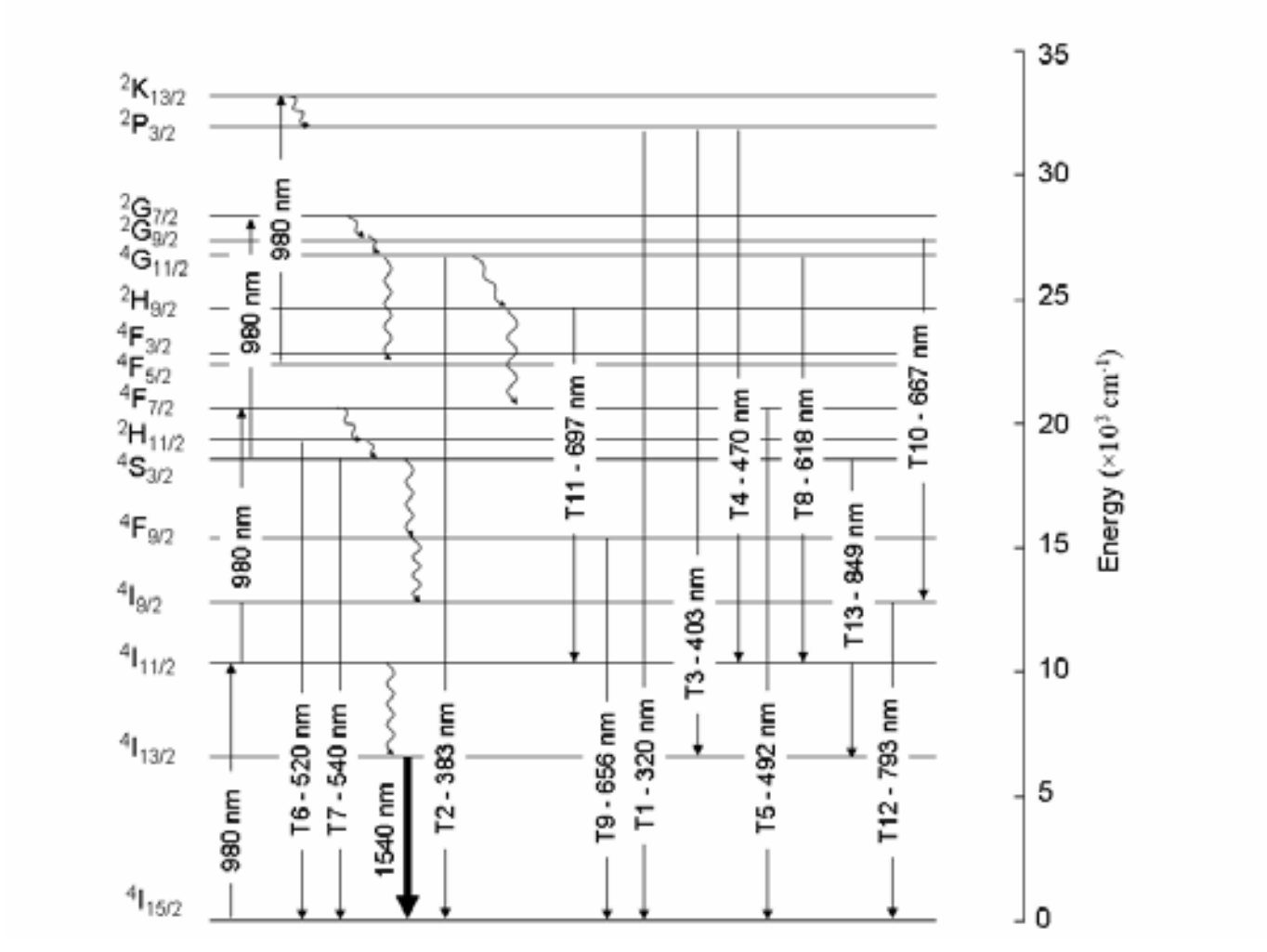

**Figure 5.** Erbium energy level diagram with radiative (solid lines) and non-radiative transitions. The thick line indicates the C-band lasing transition.

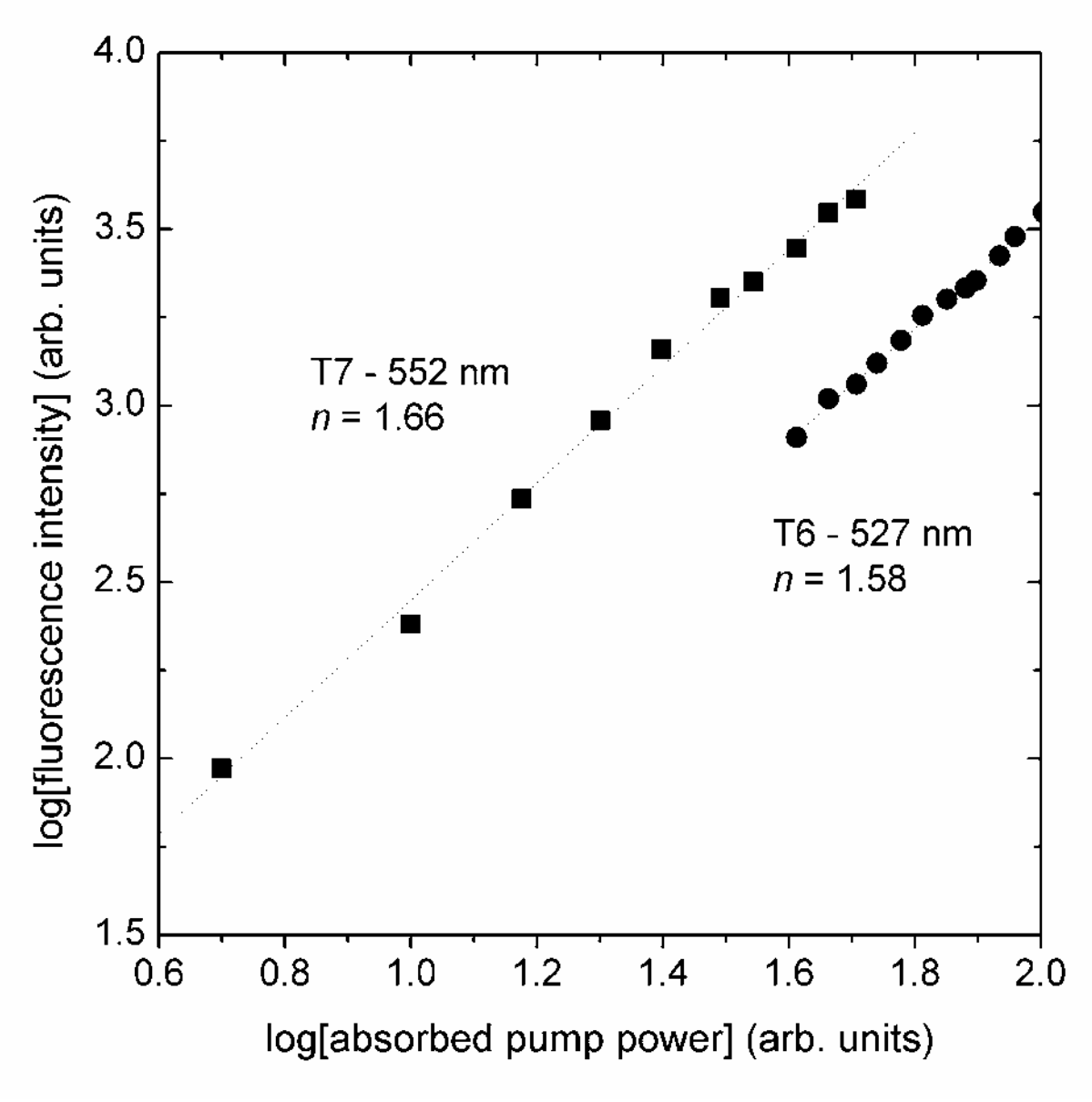

**Figure 6.** Intensity dependence of 66 $\mu$m diameter $Er^{3+}$:ZBLALiP microsphere green emissions T6 and T7 .

**Table 1.** Predicted radiative transition probabilities, electric dipole line strengths, branching ratios, and radiative lifetimes of $Er^{3+}$:ZBLALiP.

| *Transition* | *Average wavelength (nm)* | $[U^{(2)}]^2$ | $[U^{(4)}]^2$ | $[U^{(6)}]^2$ | $A_{JJ'}^{ed}$ $(s^{-1})$ | $A_{JJ'}^{md}$ $(s^{-1})$ | $S^{ed}$ $(\times 10^{-19}$ $cm^3)$ | $\beta$ | $\tau_R$ *(ms)* |
|---|---|---|---|---|---|---|---|---|---|
| $^4I_{13/2} \rightarrow {}^4I_{15/2}$ | 1519 | 0.0195 | 0.1173 | 1.4316 | 77.8 | 32.2 | 1.88 | 1.00 | 9.10 |
| $^4I_{13/2} \rightarrow {}^4I_{15/2}$ | 975 | 0.0282 | 0.0003 | 0.3953 | 97.7 | | 0.54 | 0.84 | 8.57 |
| $\rightarrow {}^4I_{13/2}$ | 2727 | 0.0210 | 0.1100 | 1.0400 | 11.7 | 7.2 | 1.43 | 0.16 | |
| $^4I_{9/2} \rightarrow {}^4I_{15/2}$ | 793 | 0.0000 | 0.1733 | 0.0099 | 105.9 | | 0.27 | 0.77 | 7.23 |
| $\rightarrow {}^4I_{13/2}$ | 1668 | 0.0003 | 0.0081 | 0.6400 | 31.8 | | 0.75 | 0.23 | |
| $\rightarrow {}^4I_{11/2}$ | 4663 | 0.0030 | 0.0674 | 0.1271 | 0.5 | | 0.26 | 0.00 | |
| $^4F_{9/2} \rightarrow {}^4I_{15/2}$ | 656 | 0.0000 | 0.5354 | 0.4618 | 946.5 | | 1.33 | 0.89 | 0.939 |
| $\rightarrow {}^4I_{13/2}$ | 1025 | 0.0096 | 0.1576 | 0.0870 | 66.8 | | 0.36 | 0.06 | |
| $\rightarrow {}^4I_{11/2}$ | 1966 | 0.0671 | 0.0088 | 1.2611 | 43.5 | 5.0 | 1.67 | 0.04 | |
| $\rightarrow {}^4I_{9/2}$ | 3623 | 0.0960 | 0.0061 | 0.0120 | 1.3 | 2.2 | 0.32 | 0.00 | |
| $^4S_{3/2} \rightarrow {}^4I_{15/2}$ | 540 | 0.0000 | 0.0000 | 0.2211 | 699.5 | | 0.25 | 0.66 | 0.949 |
| $\rightarrow {}^4I_{13/2}$ | 849 | 0.0000 | 0.0000 | 0.3481 | 294.3 | | 0.40 | 0.28 | 2.823 |
| $\rightarrow {}^4I_{11/2}$ | 1212 | 0.0000 | 0.0037 | 0.0789 | 23.5 | | 0.10 | 0.02 | |
| $\rightarrow {}^4I_{9/2}$ | 1688 | 0.0000 | 0.0729 | 0.2560 | 36.4 | | 0.40 | 0.03 | |
| $^2H_{11/2} \rightarrow {}^4I_{15/2}$ | 520 | 0.7056 | 0.4109 | 0.0870 | 3453.2 | | 2.91 | 0.92 | 0.265 |
| $\rightarrow {}^4I_{13/2}$ | 792 | 0.0230 | 0.0611 | 0.0527 | 75.2 | 54.2 | 0.22 | 0.03 | |
| $\rightarrow {}^4I_{11/2}$ | 1115 | 0.0357 | 0.1382 | 0.0371 | 43.3 | 81.1 | 0.36 | 0.03 | |
| $\rightarrow {}^4I_{9/2}$ | 1507 | 0.2077 | 0.0662 | 0.2858 | 52.5 | 0.0 | 1.08 | 0.01 | |
| $\rightarrow {}^4F_{9/2}$ | 2579 | 0.3629 | 0.0224 | 0.0022 | 11.4 | 0.2 | 1.17 | 0.00 | |
| $^4F_{7/2} \rightarrow {}^4I_{15/2}$ | 492 | 0.0000 | 0.1467 | 0.6273 | 1964.4 | | 0.94 | 0.78 | 0.397 |
| $\rightarrow {}^4I_{13/2}$ | 727 | 0.0000 | 0.3371 | 0.0001 | 316.0 | | 0.50 | 0.13 | |
| $\rightarrow {}^4I_{11/2}$ | 983 | 0.0035 | 0.2648 | 0.1515 | 147.2 | | 0.58 | 0.06 | |
| $\rightarrow {}^4I_{9/2}$ | 1245 | 0.0163 | 0.0954 | 0.4277 | 85.5 | | 0.68 | 0.03 | |
| $\rightarrow {}^4F_{9/2}$ | 1947 | 0.0121 | 0.0342 | 0.0151 | 3.5 | | 0.11 | 0.00 | |
| $^4F_{5/2} \rightarrow {}^4I_{15/2}$ | 450 | 0.0000 | 0.0000 | 0.2237 | 873.1 | | 0.26 | 1.00 | 1.15 |
| $^4F_{3/2} \rightarrow {}^4I_{15/2}$ | 443 | 0.0000 | 0.0000 | 0.1204 | 690.2 | | 0.14 | 1.00 | 1.44 |
| $^2H_{9/2} \rightarrow {}^4I_{15/2}$ | 409 | 0.0000 | 0.0190 | 0.2255 | 831.1 | | 0.29 | 0.33 | 0.400 |
| $\rightarrow {}^4I_{13/2}$ | 556 | 0.0780 | 0.1194 | 0.3535 | 952.7 | | 0.83 | 0.38 | |
| $\rightarrow {}^4I_{11/2}$ | 697 | 0.0428 | 0.0824 | 0.1128 | 225.9 | 31.0 | 0.39 | 0.10 | |
| $\rightarrow {}^4I_{9/2}$ | 823 | 0.0147 | 0.0062 | 0.0043 | 21.2 | 0.9 | 0.06 | 0.01 | |
| $\rightarrow {}^4F_{9/2}$ | 1079 | 0.0055 | 0.0314 | 0.0369 | 183.8 | 36.3 | 0.11 | 0.09 | |
| $\rightarrow {}^2H_{11/2}$ | 1854 | 0.0308 | 0.1828 | 0.0671 | 151.9 | 0.9 | 0.45 | 0.06 | |
| $\rightarrow {}^4F_{7/2}$ | 2485 | 0.1058 | 0.0488 | 0.0240 | 60.9 | 0.7 | 0.43 | 0.02 | |
| $^4G_{11/2} \rightarrow {}^4I_{15/2}$ | 383 | 0.9178 | 0.5271 | 0.1197 | 11619. | | 3.79 | 0.83 | 0.072 |
| $\rightarrow {}^4I_{13/2}$ | 505 | 0.1011 | 0.2642 | 0.2550 | 8 | 37.7 | 1.00 | 0.10 | |
| $\rightarrow {}^4I_{11/2}$ | 618 | 0.0002 | 0.4930 | 0.0144 | 1300.0 | 0.1 | 0.75 | 0.04 | 0.989 |
| $\rightarrow {}^4I_{9/2}$ | 724 | 0.0645 | 0.0117 | 0.0467 | 527.4 | 0.6 | 0.27 | 0.01 | |
| $\rightarrow {}^4F_{9/2}$ | 905 | 0.4436 | 0.0388 | 0.0104 | 119.8 | 2.6 | 1.45 | 0.02 | |
| $\rightarrow {}^2H_{11/2}$ | 1394 | 0.0006 | 0.1600 | 0.1100 | 327.5 | 10.2 | 0.37 | 0.00 | |
| $^4G_{9/2} \rightarrow {}^4I_{15/2}$ | 365 | 0.0000 | 0.2416 | 0.1235 | 2033.6 | | 0.50 | 1.00 | 0.492 |
| $^2G_{7/2} \rightarrow {}^4I_{15/2}$ | 358 | 0.0000 | 0.0174 | 0.1163 | 838.9 | | 0.16 | 0.95 | 1.132 |
| $^2P_{3/2} \rightarrow {}^4I_{15/2}$ | 320 | 0.0000 | 0.0000 | 0.0260 | 402.7 | | 0.03 | 0.12 | 0.310 |
| $\rightarrow {}^4I_{13/2}$ | 403 | 0.0000 | 0.0000 | 0.1600 | 1241.6 | | 0.18 | 0.38 | 0.353 |
| $\rightarrow {}^4I_{11/2}$ | 470 | 0.0000 | 0.1300 | 0.0250 | 936.3 | | 0.22 | 0.29 | 0.629 |
| $\rightarrow {}^4I_{9/2}$ | 524 | 0.0000 | 0.0440 | 0.0092 | 230.4 | | 0.08 | 0.07 | |
| $\rightarrow {}^4F_{9/2}$ | 616 | 0.0000 | 0.0560 | 0.0045 | 164.3 | | 0.09 | 0.05 | |
| $\rightarrow {}^4S_{3/2}$ | 763 | 0.0847 | 0.0000 | 0.0000 | 258.8 | | 0.26 | 0.08 | |
| $^2K_{13/2} \rightarrow {}^4I_{15/2}$ | 302 | 0.0032 | 0.0029 | 0.0152 | 167.0 | | 0.03 | 1.00 | 5.99 |